\documentclass[10pt,conference]{IEEEtran}
\IEEEoverridecommandlockouts
\usepackage{cite}
\usepackage{amsmath,amssymb,amsfonts}
\usepackage{algorithmic}
\usepackage{graphicx}
\usepackage{textcomp}
\usepackage{xcolor}
\usepackage{xspace}
\usepackage{booktabs}
\usepackage{balance}
\usepackage[T1]{fontenc}

\usepackage{array}
\usepackage{multirow}
\usepackage{enumitem}
\usepackage{dashrule}
\usepackage{float}
\usepackage{pifont}
\usepackage{cleveref}
\usepackage{colortbl}
\usepackage{amsthm}
\usepackage{tabulary}
\usepackage{lipsum}
\usepackage{tcolorbox}
\tcbuselibrary{breakable}
\usepackage{makecell}

\newcommand{\tool}{ccenv\xspace}

\newcommand{\Fig}{Fig.\xspace}
\newcommand{\Tab}{Table\xspace}
\newcommand{\Sec}{Section}
\newcommand{\ccpp}{C/C++\xspace}

\newcommand{\saveSpaceFig}{\vspace{-6pt}}

\DeclareRobustCommand{\mybox}[2][gray!20]{%
\begin{tcolorbox}[
        breakable,
        left=0pt,
        right=0pt,
        top=0pt,
        bottom=0pt,
        colback=#1,
        colframe=#1,
        width=\linewidth, 
        enlarge left by=0mm,
        boxsep=5pt,
        arc=0pt,outer arc=0pt,
        ]
        #2
\end{tcolorbox}
}

\usepackage{url}
\makeatletter
\newcommand{\linebreakand}{
  \end{@IEEEauthorhalign}
  \hfill\mbox{}\par
  \mbox{}\hfill\begin{@IEEEauthorhalign}
}
\def\UrlAlphabet{%
 \do\a\do\b\do\c\do\d\do\e\do\f\do\g\do\h\do\i\do\j%
 \do\k\do\l\do\m\do\n\do\o\do\p\do\q\do\r\do\s\do\t%
 \do\u\do\v\do\w\do\x\do\y\do\z\do\A\do\B\do\C\do\D%
 \do\E\do\F\do\G\do\H\do\I\do\J\do\K\do\L\do\M\do\N%
 \do\O\do\P\do\Q\do\R\do\S\do\T\do\U\do\V\do\W\do\X%
 \do\Y\do\Z}
\def\UrlDigits{\do\1\do\2\do\3\do\4\do\5\do\6\do\7\do\8\do\9\do\0}
\g@addto@macro{\UrlBreaks}{\UrlOrds}
\g@addto@macro{\UrlBreaks}{\UrlAlphabet}
\g@addto@macro{\UrlBreaks}{\UrlDigits}

\makeatother

\long\def\com#1{}

\begin{document}

\title{Uncovering and Mitigating the Impact of Frozen Package Versions for Fixed-Release Linux}

\author{\IEEEauthorblockN{Wei Tang}
\IEEEauthorblockA{\textit{School of Software} \\
\textit{Tsinghua University}\\
Beijing, China\\
tangw13\_thu@163.com}
\and
\IEEEauthorblockN{Zhengzi Xu}
\IEEEauthorblockA{\textit{School of Computer Science and Engineering} \\
\textit{Nanyang Technological University}\\
Singapore\\
zhengzi.xu@ntu.edu.sg}
\and
\IEEEauthorblockN{Chengwei Liu}
\IEEEauthorblockA{\textit{School of Computer Science and Engineering} \\
\textit{Nanyang Technological University}\\
Singapore\\
chengwei001@e.ntu.edu.sg}
\linebreakand
\IEEEauthorblockN{Ping Luo}
\IEEEauthorblockA{\textit{School of Software} \\
\textit{Tsinghua University}\\
Beijing, China\\
luop@mail.tsinghua.edu.cn}
\and
\IEEEauthorblockN{Yang Liu}
\IEEEauthorblockA{\textit{School of Computer Science and Engineering} \\
\textit{Nanyang Technological University}\\
Singapore\\
yangliu@ntu.edu.sg}
}

\maketitle
\thispagestyle{plain}
\pagestyle{plain}

\begin{abstract}
Fixed-release Linux distributions like Debian have gained widespread adoption due to their exceptional stability and internal compatibility. However, software evolves over time, and disparities arise and continue to accumulate across different releases of one Linux distribution, making it increasingly challenging to port software and fix security issues for different releases. Such disparities could be amplified by software dependencies and impact the whole ecosystem. Existing research only considers internal characteristics within one release and neglects compatibility issues and security threats that are caused by the ecosystem gap between releases. Besides, existing package managers that can bridge the ecosystem gap rely on designing a new tool with self-built central repository rather than reusing Debian mirrors to improve the built-in package manager.


Towards understanding the ecosystem gap of fixed-release Linux that is caused by the evolution of mirrors, we conducted a comprehensive study of the Debian ecosystem. This study involved the collection of Debian packages and the construction of the dependency graph of the Debian ecosystem. Utilizing historic snapshots of Debian mirrors, we were able to recover the evolution of the dependency graph for all Debian releases, including obsolete ones. Through the analysis of the dependency graph and its evolution, we investigated from two key aspects: (1) compatibility issues and (2) security threats in the Debian ecosystem. Our findings provide valuable insights into the use and design of Linux package managers. To address the challenges revealed in the empirical study and bridge the ecosystem gap between releases, we propose a novel package management approach allowing for separate dependency environments based on native Debian mirrors. We present a working prototype, named \tool, which can effectively remedy the inadequacy of current tools.
\end{abstract}


\section{Introduction}\label{sec:intro}
Fixed-release Linux distributions, such as Debian, Ubuntu, and Fedora, have become the foundation of the entire cyberspace and powers critical services in the fields of cloud computing, network server, containerization, embedded devices~\cite{cozzi2018understanding}, etc. Mirrors of fixed-release Linux distributions are released regularly every few months. Only the security updates are released frequently, and new software packages or updates to existing software are held back until a new version of the operating system is released after a fixed period of time.

The update policy of fixed-release model ensures high levels of system reliability and internal compatibility, however it causes the old releases to lag continuously and ecosystem gap forms between releases. Disparities of software ecosystems continue to accumulate across different releases, making it increasingly challenging to port software and fix security issues. Such disparities could be amplified by software dependencies and then impact the whole ecosystem. There are two kinds of problems caused by the fixed release model, incompatibility and security threats. As the maintainer would not update released packages, package lag leads to incompatibility of the system environment with the subsequent dependency requirements. Additionally, the fixed-release model leads to version lag in the deployment of updated versions with vulnerability fixes, resulting in a persistent accumulation of security threats over time.

The following examples illustrate the compatibility issues and security threats between releases for fixed-release model. In \verb|Ubuntu-devel-discuss| archives, many users have requested new package integration to improve the usability, since the lack of some specific package affects the usability of the system in practice. For instance, Tomcat SSL Connector breaks under Linux in Tomcat 9.0.31~\cite{url:tomcat_issue}, and developers have requested an update to the official Ubuntu 20.04 repositories to include a new version of Tomcat~\cite{url:lack_version1}.
However, updating Tomcat version is hindered by the \verb|StableReleaseUpdates| policy~\cite{sru:online} of fixed release model, which allows for updates only in the case of high-impact security issues or safe updates that have a low potential for regressing existing installations.
Another example is Borealis, which depends on \verb|glfw| version 3.3 or higher, causing it to only work on Ubuntu 19.10 or later~\cite{borealis:online}. For old releases such as Ubuntu 18.04, users need to manually install a newer version of \verb|glfw| package. In addition to compatibility issues, the lack of timely updates would also increase safety risks for users. For example, \verb|CVE-2021-30560|, affects \verb|libxslt| before 1.1.35~\cite{libxslt_vul:online}, and all Debian releases are affected, since they host old package versions that have not been updated to version 1.1.35 due to the \verb|StableReleaseUpdates| policy. 

The ecosystem gap between releases impacts the compatibility and security of the system. It is imperative to perform an exploratory study on issues about dependencies in the software ecosystem for fixed-release Linux, especially those caused by the ecosystem gap between releases. Existing works on dependency analysis in the Linux ecosystem has primarily focused on static dependencies at a specific point in time~\cite{gonzalez2009macro, de2009analysis, artho2012software, wang2015graph, wang2019analysis, zerouali2021multi}. However, the limitations of this approach are clear as it lacks information on the evolution of mirrors over time, making it difficult to track the creation, propagation, repair, and impact of problems in the ecosystem across multiple releases.

Existing empirical studies on evolutionary mirrors at the ecosystem level focused on language-specific package managers such as Maven, NPM, or .NET~\cite{liu2022demystifying, wang2020empirical, li2022nufix, huang2022characterizing, chen2021helping, chowdhury2021untriviality, wang2018dependency, latendresse2022not}, using evolutionary dependency graph to analyze dependencies within the ecosystem. However, the construction of an evolutionary dependency graph for fixed-release Linux is challenging, as Linux mirrors are dynamic and constantly changing. Mirrors only provides one version of a package at any given time, making it difficult to trace historic dependencies. Due to the lack of a comprehensive and evolutionary dependency graph, none of the existing works have investigated the dependency issues in the ecosystem on fixed-release Linux, especially in terms of compatibility and security over time.

Furthermore, existing approaches, that can make up for the shortcomings of Linux built-in package managers such as Nix~\cite{nix:online} and Brew~\cite{homebrew:online}, always rely on creating a new tool and building new central repositories rather than bridging the ecosystem gap based on native Linux mirrors. Besides, creating a new software repository for the entire ecosystem is an incredibly challenging task. The central repository requires extensive manual maintenance to ensure the security and reliability~\cite{Duan2021TowardsMS, zimmermann2019small, kikas2017structure}.

\begin{figure*}[t]
    \centering
    \includegraphics[width=0.8\textwidth]{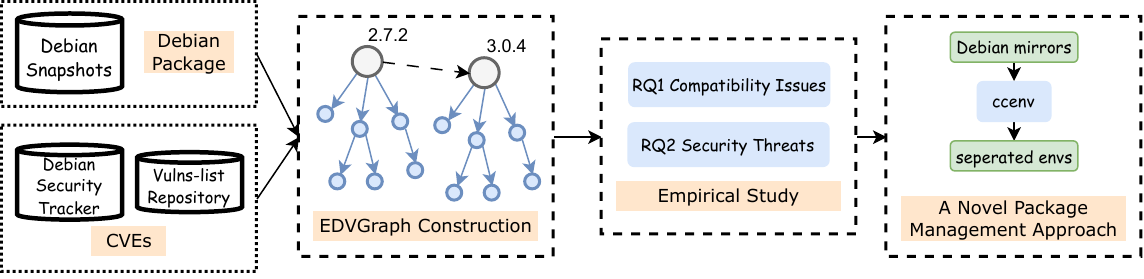}
    \saveSpaceFig
    \caption{Overview of our work.}
    \label{fig:overview_study}
    \saveSpaceFig
\end{figure*}

The overview of our work in this paper is shown in \Fig~\ref{fig:overview_study}. To address the challenge of constructing an evolutionary dependency graph, we build a framework that can automatically collect packages, parse package information, and construct the evolutionary dependency graph. Since Debian and Debian-based systems are the most popular Linux distributions~\cite{os_share:online}, we implement the framework on Debian mirrors and construct the evolutionary dependency graph (EDGraph) in the Debian ecosystem. We consider the last five Debian releases, which span over 11 years from 2011 to 2022 and include all stages in the life of Debian. Totally, we obtain over 37K projects, 102K packages and 1.5 million package versions. To investigate the security issues, the framework collects CVE(Common Vulnerabilities and Exposures) vulnerabilities from the Debian ecosystem, resolves the affected nodes in the dependency graph, then assigns CVEs to affected nodes. Based on the evolutionary dependency-vulnerability graph (EDVGraph), we conduct an empirical study from two aspects. First, we examine the compatibility issues, including project differences and version conflicts across Debian releases, in order to understand the project lag over time and the influence of the fixed-release model on compatibility across Debian releases. Second, we investigate the impact of vulnerabilities in the Debian ecosystem to assess the potential impact of fixed-release model on security. Our empirical study on the Debian ecosystem reveals many valuable findings. Based on these results, we propose a novel package management approach to bridge the ecosystem gap between releases, which creates an independent workspace for each package and resolves dependencies in an isolated environment. We implement a proof-of-concept tool, \tool and evaluate it with real-world issues. Our results demonstrate that \tool can effectively address the limitations of current tools.

In summary, our contributions are as follows:
\begin{itemize}[leftmargin=4mm]
\item We build a framework to collect Debian packages and construct the evolutionary dependency graph that is assigned with CVEs for the entire Debian ecosystem. The framework and the knowledge graph could facilitate the relevant analysis of the Linux ecosystem.

\item We conduct a large-scale empirical study on the Debian ecosystem and discuss the research questions about compatibility issues and security threats. 

\item We propose a novel package management approach and implement a proof-of-concept tool, \tool, to addresses the limitations of existing tools. Our evaluation results demonstrate that it can serve as a complementary tool to existing tools.

\item Code of the framework\footnote{https://anonymous.4open.science/r/sysdep-D6E6} for graph construction and the dependency graph\footnote{https://figshare.com/s/6d637a65cbd5387fcbd5} data is made publicly available to facilitate the research in related fields. We have open sourced \tool\footnote{https://anonymous.4open.science/r/ccenv-B7F5} to provide a useful tool for package management on Linux.
\end{itemize}

\section{Background}
\subsection{Glossary}
We explain two important concepts to prevent possible confusion.

\noindent\textbf{Package.} In Debian, there are two types of packages: binary packages and source packages. Binary packages are used for software installation, while source packages are archives of the original source code. In this paper, the term \verb|package| refers specifically to binary packages without consideration of source packages.

\noindent\textbf{Project.} Projects in Debian mirrors refer to the upstream repositories, such as GitHub repositories or open-source software. Thus, we consider the source repository name or software name as the project name. 

A project provides one or multiple packages. For example, the project \verb|FFmpeg|~\cite{ffmpeg:online} provides multiple binary packages, such as ffmpeg, libavcodec-dev, libavdevice-dev, etc. Software version is always assigned to project and all packages from the same project share the same version. Project represents the richness of the software ecosystem. In this paper, we use indicators at project level to measure the impact of fixed-release model in the Debian ecosystem.

\subsection{Fixed-Release Model}
\begin{figure}[t]
    \centering
    \includegraphics[width=0.8\columnwidth]{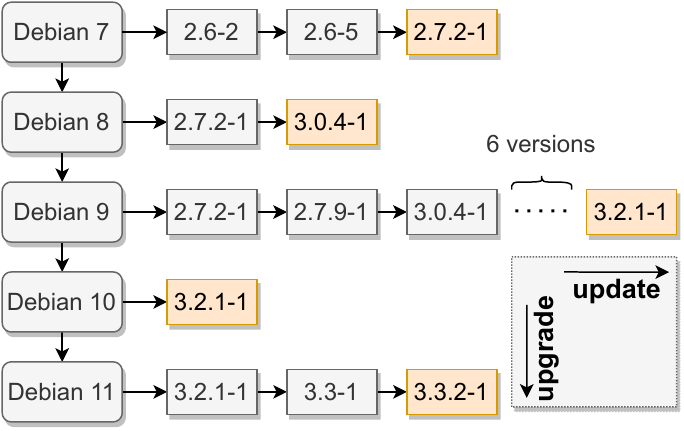}
    \saveSpaceFig
    \caption{Version evolution mechanism of Debian packages using the example of Glfw. Orange boxes present the package versions after Debian distributions are released, and gray boxes present the package updates during the time the distribution is under development, before release.}
    \saveSpaceFig
    \label{fig:update_mechanism}
\end{figure}

The fixed-release model freezes main versions of packages after they are released. \Fig~\ref{fig:update_mechanism} presents the package version evolution mechanism of Debian. We can see that there are two directions for version evolution, update and upgrade.

The update direction refers to the version evolution within the life of a Debian release, and packages are frequently updated only during the development period (i.e., gray boxes), while versions are fixed after they are released (i.e., orange boxes). The \verb|StableReleaseUpdates| policy~\cite{sru:online} specifies that released packages will not be updated unless there are high-impact bugs being reported. In the upgrade direction, the version of the Debian release evolves over time, and stable mirrors of Debian releases are created as branches from the \verb|unstable| mirror, resulting in different package versions along with the creation time.

Each release of Linux distribution with fixed-release model is only supported for a limited period. The development stage occurs before release. Once released, the Debian distribution enters its formal Debian life stage, which lasts for about three years, followed by the long-term support(LTS) stage that extends the lifespan of Debian to five years. When Debian releases reach the end of support, they would expire and no longer be maintained by official teams.

The fixed-release model is a trade-off between stability and up-to-date software, and it focuses on stability and reliability rather than on providing the latest software packages. Outdated software and limited software availability
increase package lags in old systems, which causes problems with package usage in certain scenarios.

\section{DepGraph Construction}\label{sec:method}
\begin{figure}[t]
    \centering
    \includegraphics[width=0.45\textwidth]{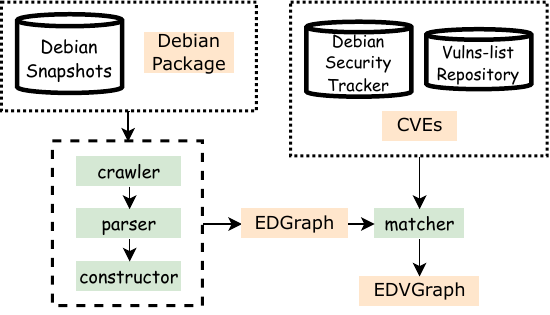}
    \saveSpaceFig
    \caption{Overview of the framework for evolutionary dependency-vulnerability graph construction.}
    \label{fig:framework}
    \saveSpaceFig
\end{figure}

We design and implement a framework to collect Debian packages and construct the evolutionary dependency-vulnerability graph to support the large-scale study on package dependencies in the Debian ecosystem. The framework, as shown in Figure~\ref{fig:framework}, consists of four modules: the crawler, the parser, the constructor and the matcher. More details about the implementation can be accessed in our source code repository.

\subsection{Crawler}
The crawler module is responsible for collecting the package index and the metadata for Debian releases. In Debian mirrors, a \verb|Packages.gz| file is created to contain all package indices and the metadata, including package name and dependency relationships. To construct the evolutionary dependency graph over time, the crawler module crawls all \verb|Packages.gz| files in the snapshots of Debian mirrors from the official archives~\cite{snapshot:online}. The snapshots of Debian mirrors are created every six hours, providing access to old packages. However, parsing the diff of six-hour intervals incurs high storage requirements and does not offer much wider benefit than a one-day interval. Thus, we collect the first snapshot of Debian mirrors for each day.

\subsection{Parser}
\begin{figure}[t]
    \centering
    \includegraphics[width=0.7\columnwidth]{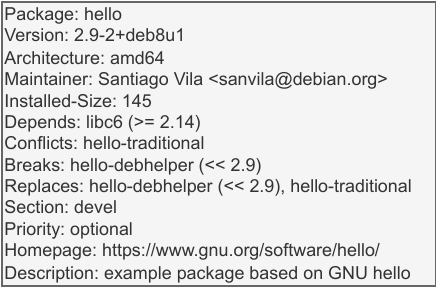}
    \saveSpaceFig
    \caption{An example of control file~\cite{debian_package:online}.}
    \saveSpaceFig
    \label{fig:control_file}
\end{figure}

We build a text parser based on the format of Debian control files~\cite{control:online} to extract package information. The Debian control file contains critical meta-information that is used by package management systems. As shown in Figure~\ref{fig:control_file}, a control file consists of a list of fields, including \verb|Package|, \verb|Version|, and \verb|Depends|. In a Debian mirror, the package name and version can be combined to uniquely identify a package. The \verb|Depends| field of a package describes the dependency relationships to other packages and version constraints of dependencies. Therefore, we extract dependencies by parsing the \verb|Depends| field to construct the dependency graph of the Debian ecosystem.

\subsection{Constructor}
The constructor module connects all packages through dependency relationships. Based on these relationships, dependencies at the project level can be inferred. We consider that Project A depends on Project B if there exists a package of Project A that depends on any package of B. Once we have parsed the dependencies of projects, we will obtain the EDGraph in the Debian ecosystem. The graph includes package and project nodes, which are attributed with meta-information, such as Debian release, survival time, etc. Dependencies at both the package level and the project level are added between nodes. The entire graph is publicly available in our database, presented in the Neo4j~\cite{neo4j:online} graph format.

\subsection{Matcher}
The vulnerability matcher module is designed to map CVEs to affected package versions and locate the fixed versions. Due to the lack of a complete CVE database in the Debian ecosystem, we collect vulnerability information from two databases, Debian security tracker~\cite{debian_security:online} and the GitHub repository, vuln-list~\cite{vuln_list:online}. The former contains information about current open security threats in Debian, and the latter provides fixed package versions for all vulnerabilities. Vulnerable points (i.e. individual CVE introductions) and fixed points (i.e. CVE fix in a package) can be located in the EDGraph by matching the unique combination of package name and version. Vulnerability information is attributed to package nodes in EDGraph to construct EDVGraph for a further empirical study on security threats.

\section{Empirical Study}\label{sec:empirical_study}
We have implemented our EDVGraph construction framework on five Debian releases and conducted an empirical study on package dependencies in the Debian ecosystem. The study aims to investigate two research questions related to compatibility issues and security threats to help understand and mitigate potential risks.

\subsection{Dataset}
We selected mirrors of the last five Debian releases (Debian 7 to 11) that span more than 11 years from 2011 to 2022, covering all stages of Debian lifecycle, including development, formal Debian life, and long-term support. We collected daily snapshots of five Debian mirrors and obtained 102K packages with 1.5 million versions from 37K projects. Eight million dependencies at the package level and 1.8 million at the project level are extracted to construct the EDGraph.

\subsection{RQ1: Compatibility issues}\label{sec:compability}
As a fixed release distribution, each Debian release has excellent stability and internal compatibility. On the contrary, there are significant disparities among packages and package versions, leading to incompatibility issues across different Debian releases especially old releases. Prior studies\cite{artho2012software, claes2015historical, zerouali2021multi} have investigated dependency conflicts and version lags in the Debian ecosystem. However, no previous work has explored the incompatibility across Debian releases.

\subsubsection{\textbf{How does the number of projects evolve across Debian releases?}}~\label{rq1-1}
As the \verb|unstable| Debian mirror is in rolling development, and Debian releases (\verb|Stable| mirrors) are created from \verb|unstable| mirror at different time, projects might be added or removed from mirrors of different releases. Thus, different Debian releases might have different projects in mirrors and uses of different releases suffer from different capacity of software supply. We first investigate the project evolution in Debian mirrors.

\begin{figure}[t]
    \centering
    \includegraphics[width=0.8\columnwidth]{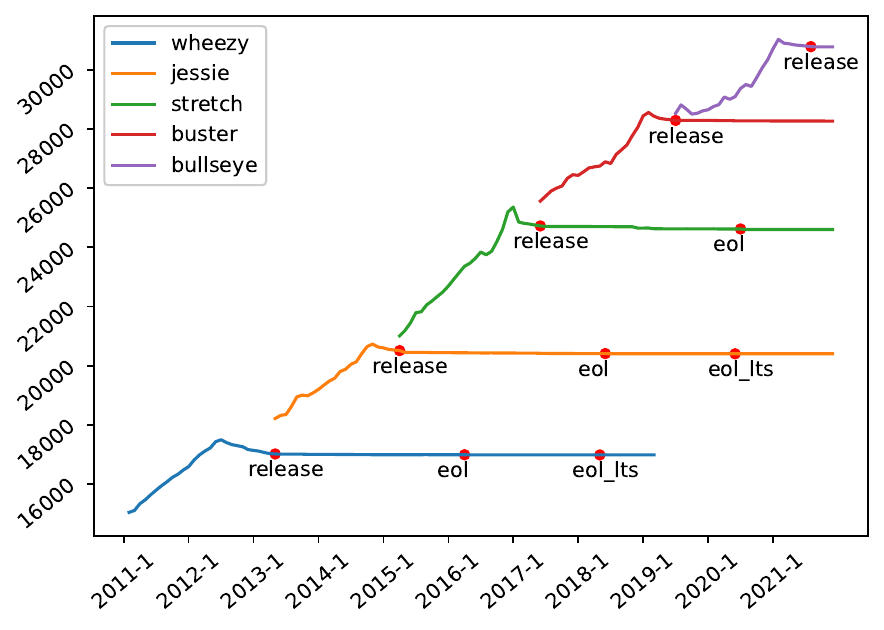}
    \saveSpaceFig
    \caption{The number of projects on Debian over time. Eol: end of life. Eol\_lts: end of long term support.}
    \label{fig:project_evolution}
\end{figure}

\begin{table}[t]
  \centering
  \caption{The project differences between five Debian releases. The element in the "wheezy" row and "jessie" column means 4752 projects exist on jessie, but not on wheezy.}
  \saveSpaceFig
  \label{tab:changes_of_projects}
  \scalebox{1}{
  \begin{tabular}{lccccc}
    \toprule
                & wheezy  & jessie & stretch & buster & bullseye        \\ 
    \midrule
        wheezy  &   /   &   4752    &   10095   &   14736   &   18210   \\
        jessie  &   1335&   /       &   5933    &   10776   &   14496   \\
        stretch &   2476&   1731    &   /       &   5486    &   9483    \\
        buster  &   3453&   2910    &   1822    &   /       &   4572    \\
        bullseye&   4419&   4122    &   3311    &   2064    &   /       \\
    \bottomrule
  \end{tabular}
}
\saveSpaceFig
\end{table}

\Fig~\ref{fig:project_evolution} presents the number of projects in five Debian releases. The number of projects shows a steady linear increase during development and then stabilizes after the release. When a Debian distribution is released, the development of the next Debian release commences, and it continues for about two years, during which a considerable number of new projects are added to the upcoming Debian release. We compared the five Debian releases and obtained the added/removed projects, as shown in \Tab~\ref{tab:changes_of_projects}. Over the 10 years from Debian 7 (wheezy) to Debian 11 (bullseye), the project number has more than doubled. A new Debian release typically adds around 4.5K to 6K new projects and removes 1.3K to 2K old projects. 12,142 projects exist in all five Debian releases. About 85\% of added/removed projects with tags are in four types of roles: shared-lib, program, devel-lib, and plugin, implying that changes primarily concentrate on software projects and affect the software supply capacity of different releases. The differences in projects can cause a lot of trouble for users working on incompatible Debian releases, as reported in a real-world case\cite{nnn:online}.

\mybox{Answer-1: \ding{172} In the development of the next Debian release, substantial new projects (4.5K to 6K) are added, and many old projects (1.3K to 2K) will be removed, which results in the differences in projects between Debian releases. \ding{173} Most project changes (85\%) concentrate on software projects, leading to different software supply capacity on different Debian releases.}

\subsubsection{\textbf{What is the span of version updates in the life of Debian releases?}}
\begin{figure}[t]
    \centering
    \includegraphics[width=0.8\columnwidth]{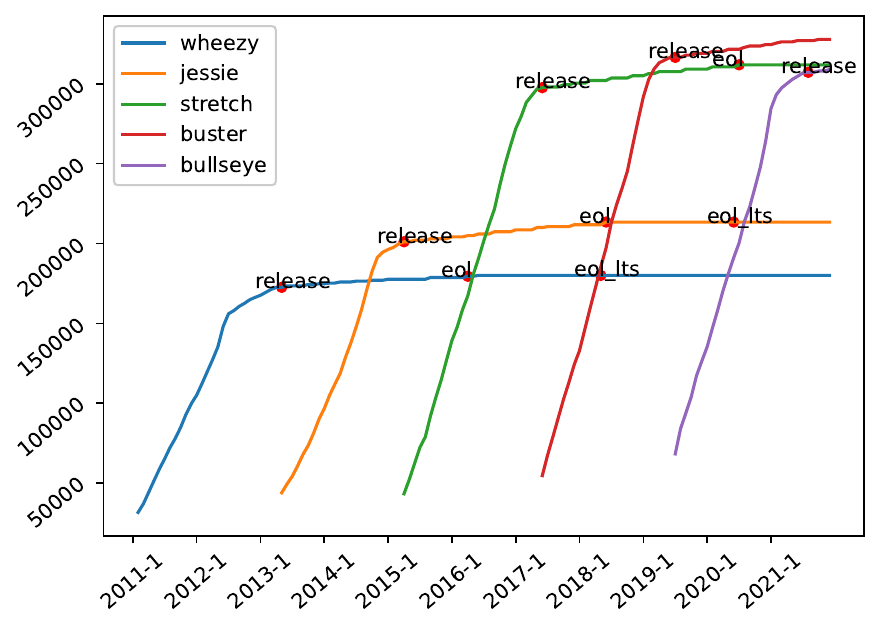}
    \saveSpaceFig
    \caption{The number of accumulated updates on Debian.}
    \saveSpaceFig
    \label{fig:package_update_evolution}
\end{figure}
Except for added/removed projects, the versions of common projects evolve between Debian releases. We identify all version updates in the evolutionary graph and check the update date. The number of accumulated updates from the beginning of the development stage is shown in \Fig~\ref{fig:package_update_evolution}. 

Between two adjacent Debian releases, a significant percentage of common projects (70\% to 85\%) are updated to new versions. The majority of version updates (96\%) occur during the development stage, before the release of a new version of Debian. Only a small proportion of updates (4\%) occur during the formal Debian life, and less than 0.1\% occur during the period of long-term support.

To measure the extent of version changes, we analyze the semantic versioning~\cite{semantic_versions:online} scheme that is widely used in software management. This scheme consists of three version fields, in the format of MAJOR.MINOR.PATCH. The project version may be updated from $\langle major, minor, patch \rangle$ to $\langle major1, minor1, patch1 \rangle$. The span of version increment for the MAJOR field can be calculated as $major1-major$. From the start of development stage to the end of LTS, the span of version number increments on average are 0.15, 0.89, 1.49 for MAJOR, MINOR, and PATCH version, respectively. Once Debian is released, the increments decrease to 0.006, 0.034, and 0.079, respectively. These results prove that package versions remain relatively stable throughout the entire support life after release, with most changes occurring during the development period.

According to the semantic versioning scheme, a MAJOR version increment indicates backward-incompatible API changes, while compatible API additions or changes result in a MINOR version increment, and bug fixes that do not affect the API increment the PATCH version. Furthermore, a previous study~\cite{zhang2022has} has shown that even MINOR updates can introduce a high percentage of changes that are not compatible. Therefore, we can infer that incompatible version constraints are common across different releases of Debian. 

\mybox{Answer-2: \ding{172} Between two consecutive releases of Debian, a majority of common projects (70\% to 85\%) are updated to new versions. \ding{173} Significant changes are made to package versions during the development stage, resulting in incompatibility across Debian releases. The span of semantic version changes is 0.15, 0.89, and 1.49 for the MAJOR, MINOR, and PATCH fields, respectively, which causes incompatibilities across Debian releases.}

\subsubsection{\textbf{How many version conflicts are there across Debian releases?}}
As mentioned above, significant version changes are made in the development of new Debian releases. Version disparities continue to accumulate in the upgrade direction from Debian 7 to Debian 11. Debian releases may not satisfy the required dependencies by newer or older package versions like the case of Tomcat in ~\Sec~\ref{motivation}. Excellent internal compatibility sacrifices external compatibility and restricts users to package version selection. In this section, we investigate the version conflicts across Debian releases to obtain empirical understandings on incompatibility issues about package version selection.

To explore the prevalence and impact of version conflicts and incompatibilities across Debian releases, we conducted an comprehensive analysis on package dependencies in the Debian ecosystem. Our investigation focused on identifying cases where packages from one release could not be installed on another release due to incompatible dependencies.

\begin{table}[t]
  \centering
  \caption{The number of incompatible projects across Debian releases. The element in the "wheezy" row and "jessie" column means 5399 projects on jessie have incompatible version constraints that are not satisfied by projects on wheezy.}
  \saveSpaceFig
  \label{tab:incompatible_projects}
  \scalebox{1}{
  \begin{tabular}{lccccc}
    \toprule
                & wheezy  & jessie & stretch & buster & bullseye \\
    \midrule
        
        wheezy  & /     &   5399&   7174    &   7297    &   7457    \\
        jessie  & 983   &   /   &   4392    &   5127    &   5801    \\
        stretch & 2514  &   1011&   /       &   3515    &   4908    \\
        buster  & 3693  &   2981&   1131    &   /       &   3852    \\
        bullseye& 4119  &   3465&   2465    &   759     &   /       \\
    \bottomrule
  \end{tabular}
}
\end{table}


In the dependency graph, we have extracted the dependencies at the package level and project level. Specifically, we identify packages that are deemed incompatible between two Debian releases, $\langle A, B\rangle$, if the dependencies of the package on release B are not satisfied by the packages on release A. Furthermore, we identify the project that provides the incompatible package as an incompatible project of $\langle A, B\rangle$. We only consider 12,142 common projects that exist on all five Debian releases, and thus added/removed projects in \Tab~\ref{tab:changes_of_projects} are not included in the number of incompatible projects presented in \Tab~\ref{tab:incompatible_projects}. Our results show that there are more incompatible projects as the gap becomes wider between Debian versions, and the number of $\langle A, B\rangle$ is significantly larger than the number of $\langle B, A\rangle$ when Debian B is later than Debian A. This indicates that backward compatibility is better than forward compatibility in the Debian ecosystem and older Debian releases face difficulties in meeting the requirements of updating to the latest package versions.

Incompatible projects of $\langle A, B\rangle$ are caused by the inability to find the proper libraries on Debian A, where libraries should meet the version constraints defined on Debian B. In version constraints on Debian, there are five relations: $<<$, $<=$, $=$, $>=$, and $>>$, which represent strictly earlier, earlier or equal, exactly equal, later or equal, and strictly later, respectively. We have summarized four types of reasons for incompatible projects of $\langle A, B\rangle$. The four types of reasons are: (1) the host project on Debian B depends on a later or strictly later package version, but the package version on Debian A is earlier than it, (2) the host project depends on a strictly earlier or earlier package version, but the package on Debian A has a later version, (3) the host project depends on a specific version, but the package version on Debian A is not equal to it, and (4) the host project depends on a package that does not exist on Debian A. Four types of reasons are represented by $>$, $<$, $=$, and \verb|no|, respectively. We count the number of project-level version conflicts for each type, that cause incompatible projects of $\langle A, B\rangle$. The statistics of reasons is presented in \Tab~\ref{tab:imcompatible_reasons}. For the forward incompatibility of $\langle A, B\rangle$ where release B is later than A, there are two main reasons, relying on later package versions or the non-exist packages on Debian A, that contribute more than 99\% incompatible projects.

However, in the opposite case, where Debian B is earlier than Debian A in the left corner of \Tab~\ref{tab:imcompatible_reasons}, the impact of version constraints on earlier package versions increases, but it is still not comparable with non-exist packages that contribute 97\% incompatible projects. We inspect non-exist packages and find that many packages are named with version numbers, resulting in the same package having different names for different versions, such as \verb|libboost-iostreams1.55.0|~\cite{url:boost}, which can be integrated with the version constraints of "$=$". We recommend that maintainers should unify the name of the same package and project for better package indexing and management.

\begin{table}[t]
  \centering
  \caption{The number of reasons for project incompatibility across Debian releases. The element in the "wheezy" row and "jessie" column means how many incompatible projects of $\langle wheezy, jessie\rangle$ are caused by each type of reason.}
  \saveSpaceFig
  \label{tab:imcompatible_reasons}
  \scalebox{0.88}{
  \begin{tabular}{lccccc}
    \toprule
                & wheezy  & jessie & stretch & buster & bullseye \\
    \midrule
    wheezy&/&\makecell[c]{$>$:4898\\$<$:1\\=:4\\no:2821}&\makecell[c]{$>$:6433\\$<$:3\\=:3\\no:4323}&\makecell[c]{$>$:6355\\$<$:2\\=:1\\no:4564}&\makecell[c]{$>$:6350\\$<$:2\\=:0\\no:5095} \\
    \cmidrule(r){2-6}
    jessie&\makecell[c]{$>$:3\\$<$:3\\=:4\\no:977}&/&\makecell[c]{$>$:3032\\$<$:7\\=:3\\no:3115}&\makecell[c]{$>$:3660\\$<$:3\\=:1\\no:3523}&\makecell[c]{$>$:4322\\$<$:5\\=:0\\no:4246} \\
    \cmidrule(r){2-6}
    stretch&\makecell[c]{$>$:0\\$<$:94\\=:18\\no:2457}&\makecell[c]{$>$:5\\$<$:5\\=:0\\no:1007}&/&\makecell[c]{$>$:2230\\$<$:6\\=:1\\no:2129}&\makecell[c]{$>$:3430\\$<$:6\\=:4\\no:3426} \\
    \cmidrule(r){2-6}
    buster&\makecell[c]{$>$:0\\$<$:95\\=:17\\no:3648}&\makecell[c]{$>$:0\\$<$:118\\=:5\\no:2922}&\makecell[c]{$>$:3\\$<$:4\\=:2\\no:1126}&/&\makecell[c]{$>$:2636\\$<$:5\\=:4\\no:2776} \\
    \cmidrule(r){2-6}
    bullseye&\makecell[c]{$>$:0\\$<$:96\\=:16\\no:4081}&\makecell[c]{$>$:0\\$<$:121\\=:4\\no:3418}&\makecell[c]{$>$:2\\$<$:54\\=:4\\no:2447}&\makecell[c]{$>$:1\\$<$:3\\=:1\\no:756}&/ \\
    \bottomrule
  \end{tabular}
}
\end{table}

\mybox{Answer-3: \ding{172} More than half of common projects are incompatible, and the disparities in Debian versions aggravate the incompatibilities across Debian releases. \ding{173} Backward compatibility performs better than forward compatibility in the Debian ecosystem, with doubling to fivefold differences between two scenarios. \ding{174} More than 99\% of incompatible dependencies are caused by version constraints on later package versions and non-exist packages for forward incompatibility. Depending on non-exist packages is also the main reason for backward incompatibility, and 97\% of incompatibilities are caused by it.}

\subsection{RQ2: Security threats}
The goal of this section is to investigate the impact of vulnerabilities in the Debian ecosystem. Existing studies mainly focus on language-specific ecosystems like Maven and NPM. No previous work has studied the security threats in Debian ecosystems, especially through the dependency graph.

\subsubsection{\textbf{How many vulnerabilities exist in Debian releases?}}
Projects in Debian mirrors are composed of the downstream branches of the original repositories. These projects are maintained by official developers in the Debian community. Therefore, the Debian team is responsible for the security of the Debian ecosystem. It tracks security issues and responds to issues to fix the bug by patching the vulnerable version when the original repositories release a new patch version. We count the number of fixed vulnerabilities and open security issues as shown in \Tab~\ref{tab:fixed_open_vuls}.

We can see that open issues continue to accumulate in the Debian life and there are still thousands of open CVEs in the latest versions of packages. When stretch comes to the end of long term support (June 30, 2022), there are still more than five thousand open issues that are not fixed. Debian team only tracks open security issues on the latest three releases. Therefore, it is reasonable to infer that there are more open threats on outdated Debian releases (i.e. wheezy and jessie). Results prove the effectiveness and necessity of updating package versions and upgrading to new Debian releases. However, developers do not usually react in time, especially in the production environment. Furthermore, we notice that 60\% of vulnerable points are fixed in the development period. However, once released, the Debian team would have to manually patch the released version rather than update to the upstream patched version due to the fixed release model. Substantial human resources are required to fix 40\% of vulnerable points in Debian life.

Moreover, we further explore the severity of vulnerabilities~\cite{url:severity}. The statistics indicate that 1.1\%, 4.1\%, and 11.4\% of vulnerabilities have high, medium, and low security severity, respectively. It is gratifying that all vulnerabilities assigned with high or medium severity are fixed. However, we notice that 77.8\% of vulnerabilities are not assigned with severity.

\mybox{Answer-4: \ding{172} All serious vulnerabilities could be fixed in a timely manner. It is critical for users to update the package versions or upgrade to new Debian releases since the security risks continue to accumulate, and the entire Debian life after release might be affected by more than ten thousand vulnerabilities.}


\begin{table}[t]
  \centering
  \caption{open and fixed CVEs.}
  \saveSpaceFig
  \label{tab:fixed_open_vuls}
  \scalebox{1}{
  \begin{tabular}{llcccccc}
    \toprule
        Status & wheezy  & jessie  & stretch & buster & bullsey & unstable\\
    \midrule
        Open & / & / & 5427 & 4052 & 2187 & / \\
        Fixed& 5451 & 6260& 5643 & 3314& 847 & 30980\\
    \bottomrule
  \end{tabular}
}
\end{table}

\subsubsection{\textbf{How do vulnerabilities propagate in dependency graph?}}
In the dependency graph, a package has six external dependencies on average. Therefore, security threats would be amplified through direct and transitive dependencies, thus propagate in the whole Debian ecosystem.

\begin{table}[t]
  \centering
  \caption{The number of affected projects. Original: \# projects that introduce CVEs. Transitive: \# projects that are affected by transitive dependencies.}
  \saveSpaceFig
  \label{tab:affected_projects}
  \scalebox{1}{
  \begin{tabular}{lcccc}
    \toprule
        \multirow{2}{*}{Debian} & \multicolumn{2}{c}{Open} & \multicolumn{2}{c}{Fixed} \\
         & Original  & Transitive  & Original & Transitive \\
    \midrule
        wheezy  &  /  &  /  &  651 (3\%)  &  15920 (93\%)   \\
        jessie  &  /  &  /  &  733 (3\%)  &  19184 (94\%)   \\
        stretch  &  1025 (4\%)  &  23710 (96\%)  &  721 (2\%)  &  23662 (96\%)  \\
        buster  &  1018 (3\%)  &  25759 (91\%)  &  458 (1\%)  &  25605 (90\%)   \\
        bullseye  &  749 (2\%)  &  27325 (88\%)  &  156 (0\%)  &  26939 (87\%)  \\
    \bottomrule
  \end{tabular}
}
\end{table}

As presented in \Tab~\ref{tab:affected_projects}, only a small proportion of projects (less than 4\%) are vulnerable. However, almost the whole Debian ecosystem (more than 87\% of projects) is affected through transitive dependencies. The magnification scale is much larger than any other language-specific ecosystem such as Maven~\cite{prana2021out}, NPM~\cite{liu2022demystifying, zimmermann2019small, decan2018impact, zerouali2022impact}, and Python~\cite{ma2020impact}. Taking into account the severity of vulnerabilities, 33 and 205 projects out of 40K have vulnerabilities with high or medium severity, respectively. These projects can be considered as critical projects that may cause serious security problems to the system according to the severity rules~\cite{url:severity}. We investigate the impact of  critical projects for each Debian release as presented in \Tab~\ref{tab:vul_severity_impact}. It can be seen that these projects have an impact on the whole ecosystem and more than 87\% of projects are affected through transitive dependencies.

\begin{table}[t]
  \centering
  \caption{affected projects by critical projects.}
  \saveSpaceFig
  \label{tab:vul_severity_impact}
  \scalebox{1}{
  \begin{tabular}{lcccc}
    \toprule
        \multirow{2}{*}{Debian} & \multicolumn{2}{c}{High} & \multicolumn{2}{c}{Medium} \\
         & Original  & Transitive  & Original & Transitive \\
    \midrule
        wheezy  &  33 ($\sim$0\%)  &  15907 (93\%)  &  205 (1\%)  &  15973 (94\%) \\
        jessie  &  33 ($\sim$0\%)  &  19134 (93\%)  &  205 (1\%)  &  19210 (94\%) \\
        stretch  &  33 ($\sim$0\%)  &  23596 (95\%)  &  205 ($\sim$0\%)  &  23672 (96\%) \\
        buster  &  33 ($\sim$0\%)  &  25552 (90\%)  &  205 ($\sim$0\%)  &  25636 (90\%) \\
        bullseye  &  33 ($\sim$0\%)  &  26944 (87\%)  &  205 ($\sim$0\%)  &  27035 (87\%) \\
    \bottomrule
  \end{tabular}
}
\end{table}

We inspect the packages with substantial transitive dependencies and find that low-level libraries on the system, such as glibc and dbus, are the main reasons for such vulnerability propagation through dependencies in the Debian ecosystem. Different from language-specific ecosystems like Maven, which provide all released versions of a project, only one package version is available to download on Debian release at any time, because the package would be installed under the system directory by default and installing multiple versions at the same time would lead to file conflicts. Therefore, a vulnerable package version would affect all projects that depend on it and the vulnerable packages can not be avoided through version selection by users.

\mybox{Answer-5: \ding{172} The impact of vulnerable packages is amplified by transitive dependencies and propagates to the entire Debian ecosystem, especially low-level libraries. \ding{173} Vulnerable packages can not be avoided by version selection since there is only one available package version. It leads to the problem that vulnerable packages would affect all dependent projects.}

\subsubsection{\textbf{How does the lag of vulnerability fix evolve in the life of Debian releases?}}
As mentioned above, packages in Debian mirrors are downstream branches of the original repositories, and all packages are maintained by Debian team. Official developers track issues and patch vulnerable versions when the original repositories release a new patch version. Extra cycles from when the upstream solution is released to when Debian package is fixed might increase the technical lag for patching.

\begin{table}[t]
  \setlength\tabcolsep{3pt}
  \centering
  \caption{Technical lag for vulnerability fixing.}
  \saveSpaceFig
  \label{tab:fix_lag}
  \scalebox{0.88}{
  \begin{tabular}{lcccccc}
    \toprule
        \multirow{2}{*}{Debian} & \multicolumn{2}{c}{Formal Debian Life} & \multicolumn{2}{c}{LTS} & \multicolumn{2}{c}{Average} \\
        & \# Fixed Points & Lag & \# Fixed Points & Lag & \# Fixed Points & Lag \\
    \midrule
        wheezy & 2026 & 87 & 1952 & 94 & 3978 & 91 \\
        jessie & 2364 & 75 & 2671 & 222 & 5035 & 153 \\
        stretch & 2414 & 117 & 2224 & 316 & 4638 & 213 \\
        buster & 2562 & 126 & / & / & 2562 & 126 \\
        bullseye & 701 & 65 & / & / & 701 & 65 \\
    \bottomrule
  \end{tabular}
}
\end{table}

We calculate the fix lag which is the number of days from the time CVE is published in NVD to when the fixed package is uploaded. The packages containing the fixes for CVEs in EDVGraph are labeled as fixed points. There are a total of 23290 fixed points with exact CVE publish time on five Debian releases after they are formally released. As a result, 27.4\% (6,376/23,290) of vulnerable points are fixed before related vulnerabilities are published. For the remaining fixed vulnerabilities, the fix lag takes 147 days on average. Debian ecosystem has a higher quality which benefits from the strict centralized control by Debian team, compared to language-specific ecosystems, like NPM~\cite{liu2022demystifying}, in which authors of each library are responsible for the maintenance.

In terms of Debian versions, different Debian releases have different lags. As presented in \Tab~\ref{tab:fix_lag}, Debian "bullseye" has the shortest fix lag (65 days) while "stretch" has the longest (213 days). Moreover, the fix lag in the stage of LTS is generally longer than the stage of formal Debian life because LTS is maintained by a group of volunteers rather than the official Debian Security Team~\cite{lts:online}. It reflects that Debian mirror maintenance is a resource-intensive task, and the quality of Debian mirrors is affected by the number of available resources. 

\mybox{Answer-6: \ding{172} It takes a time lag of 147 days on average to fix vulnerabilities in the Debian ecosystem and 27.3\% of vulnerable points are fixed before CVE publication. \ding{173} It needs a lot of human resources to maintain Debian mirrors. Once the human resources are reduced, the quality of maintenance would be lower.}

\section{Improvement on Linux package manager}\label{sec:ccenv}
Our findings in the empirical study reveal that ecosystem gaps exist between releases for fixed-release Linux and impact the compatibility and security. Existing third-party package managers which allow different versions of the same package to coexist on the same system rely on self-built central repositories. They can not bridge the ecosystem gap between releases by installing packages from native Linux mirrors.

Inspired by CDE~\cite{guo2011cde}, we propose a novel package management approach to help users install compatible components, helping them to cope with occasional and rare scenarios without destroying the good stability of the fixed release model.

\subsection{Design}

\begin{figure}[t]
    \centering
    \includegraphics[width=0.8\columnwidth]{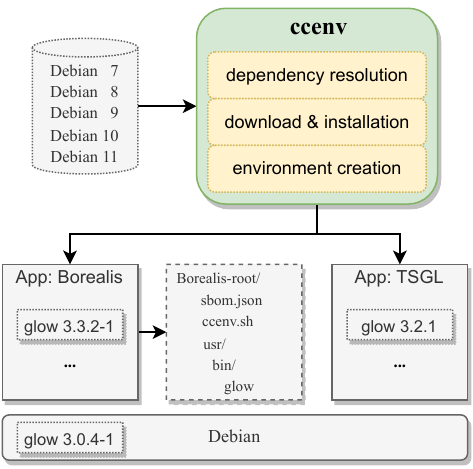}
    \saveSpaceFig
    \caption{Architecture of \tool}
    \label{fig:ccenv}
\end{figure}

In order to improve the existing Linux packages managers, some new tools are proposed such Nix~\cite{nix:online}, flatpak~\cite{flatpak:online}, Snap~\cite{snap:online}, and Homebrew on Linux~\cite{homebrew:online}. However, they create and depend on their own package repositories rather than Linux mirrors. The Linux mirrors, particularly for popular Linux distributions like Debian, are subject to stringent controls aimed at ensuring the reliability of packages and mitigating the risk of supply chain attacks.~\cite{Duan2021TowardsMS, kikas2017structure, zimmermann2019small}. It is beneficial to reuse Linux mirrors to install packages.

To overcome the limitations as mentioned above, we propose a new package management approach that retrieves packages from Debian mirrors and creates a separated workspace for individual projects. We implement a proof-of-concept tool, \tool and its architecture is shown in \Fig~\ref{fig:ccenv}. \tool first resolves all dependencies of the package which is to be installed. Then, \tool downloads all packages including host applications and required dependencies from Debian mirrors that are not limited to the same version with the local system. \tool decompresses all archived packages for installation. Finally, \tool creates the separated environment, starting entry, and SBOM files.


There are three main modules in \tool, dependency resolution, download\&installation, and environment creation. 

\noindent \textbf{Dependency Resolution.} Given the package name which is to be installed, the dependency resolution module first parses the metadata of the package in Debian mirrors and obtains all dependencies including direct and transitive dependencies. \tool supports installing packages from mirrors of different Debian releases. Resolved dependencies are correct and complete even if the package is not compatible with the system in use. As presented in \Fig~\ref{fig:ccenv}, we obtain all dependencies of \verb|Borealis|, including the item "$glow >=3.3$". The Debian release in use provides an incompatible package of \verb|glow| with version 3.0.4-1, which is ignored by \tool.

\noindent \textbf{Download \& Installation.} After all dependencies are obtained, \tool downloads relied packages from Debian mirrors. Debian binary packages are in the format of Debian-specific archive files that contain executables and some necessary files like configuration files. To install these packages, all package files are directly extracted to an independent and separated root directory, \verb|Borealis|. Contents in packages are organized through the structure of Linux directories. All extracted contents would be organized with the same directory structure under the root directory, \verb|Borealis|, not the root directory of the system ("/").

\noindent \textbf{Environment Creation.} It is critical to create an environment to make applications find dependencies and run correctly. External dependencies are defined in binaries. However, different from text files, binary files cannot be easily modified to set dependencies. We create a bash file to set environment variables such LD\_LIBRARY\_PATH which specifies directory paths that the linker should search for libraries (i.e. dependencies). All environment variables only take effect in individual environments. Multiple package versions installed on the system would not cause version conflicts, such as the dependency \verb|glow| of \verb|Borealis| and \verb|TSGL|. Besides, settings of environment variables in our bash file own a higher priority than system environment variables. Therefore, dependencies under the individual directory would be found and linked first. The incompatible package installed by default on the system would not break the running of applications. As for running the application, \tool parses commands in \verb|.desktop| files and set it in the bash file.

In addition, \tool generates a SBOM file in the json format, which contains vital information such as package name and version number. All dependencies including transitive dependencies are described in the SBOM file. It is convenient to reproduce the environment or update the installed packages using the SBOM file.

By creating individual spaces for applications, \tool makes it possible to install the latest package version, even multiple package versions from the whole Debian ecosystem. We implement the prototype of \tool. More details can be found in our open source repository.

\subsection{Validation of \tool}
Note that \tool is not proposed to replace the default Linux package managers. It is designed to make up for the deficiency of the default managers in the scenarios mentioned above. Therefore, we validate \tool with real-world incompatible projects to prove its effectiveness. We first extract all incompatible projects of $\langle buster, bullseye\rangle$ and install these projects on buster using \tool. To calculate the accuracy of successful installation,  we randomly select 50 projects with GUI interface, run them and do some operations to verify the installation. As a result, all applications are successfully installed in their individual spaces with complete and correct dependencies. 37 applications can be run directly, and the remaining 13 applications require some settings like adding paths of shared resources, like icon pictures. It could be easily fixed by manual configuration. We will implement the automatic method in \tool later due to time constraints. Results show that \tool is effective to provide users some advanced features, including multiple separated environments, fixing incompatible projects, and generating SBOM files, that are not supported by existing Linux package managers.

\section{Discussion}\label{sec:dis}

\subsection{Implications}
\noindent\textbf{For package providers and maintainers.} 
The evolutionary dependency graph in this paper provides package providers and maintainers a comprehensive perspective to understand the cost and impact of updates. For example, their own packages may be well-maintained with timely updates. However, open security issues might be introduced via dependency graph. Our dependency graph helps developers check transitive dependencies to track all potential security issues.

\tool provides a possible solution to resolve dependencies cross Debian releases without relying on the system version. With just one package, all Debian releases could install the application no matter what the Debian version is. Thus, duplicated efforts are avoided to reduce the development cost. Besides, outdated packages have a larger attack surface with more vulnerabilities and open security issues continue to accumulate over time. \tool could help providers and maintainers reduce the workload to patch old package versions. If the package is vulnerable on the system, existing tools have to install the default vulnerable package even secure alternative packages are available on other Debian releases. At this time, the vulnerable dependencies would be avoided using \tool through retrieving secure packages from other Debian releases.

\noindent\textbf{For developers on Linux.}
By utilizing the evolutionary dependency graph, developers are able to assess the availability of a Linux release and subsequently select the release that best meets their requirements. Besides, the graph can help developers estimate the cost to upgrade the system version and realize the potential problems in advance that may arise after upgrading. \tool can also allow developers to proactively manage and update the dependencies in the development environment rather than passively accept what the system provides. Developers may create virtual system environments such as Docker and virtual machines when they depend on incompatible Debian releases. \tool allows them to install packages on the local system from incompatible releases. All installed files can be accessed through the local file system. Developers could be liberated from switching different virtual system environments.

\noindent\textbf{For ordinary users.} 
Based on the results of our empirical study, we recommend users keep their systems updated. If possible, it is better to upgrade the system to the latest Debian release, especially for outdated Debian releases or releases in the period of LTS. However, it is always ignored to update the system by users, especially in production environments that require excellent stability. In such scenarios, \tool could be adopted to extend the life of Debian releases and allows outdated releases to take advantage of the latest updates. Besides, since \tool installs packages under individual directories at the user level rather than system directories at the system level, \tool would not threaten the stability of the system.

\noindent\textbf{For software repository managers.} 
There are various package managers on Linux that are always bundled with software repositories like APT, Conan~\cite{url:conan}, and Homebrew~\cite{homebrew:online}. The evolutionary dependency graph provides the capability for repository managers to track the evolution at the ecosystem level. It is helpful for various tasks such as finding critical libraries, calculating the living time of vulnerabilities, evaluating the reliability of the repository, etc.

\subsection{Does ccenv threaten the stability of applications and Linux systems?}
Ccenv can help users install package versions from different Debian releases. However, it does not check the compatibility of binaries, and it should be confirmed by users. Binary files would be changed by various complication conditions, and that is why Conan~\cite{url:conan} provides dozens of duplicated packages. Nevertheless, it is reasonable to believe that binaries in the Debian ecosystem have better compatibility than the package repository of third-party package managers like Conan and Homebrew. Therefore, \tool would have only minor effects on the stability of applications. If the installed application through \tool does not run correctly, users can still install it using the default package managers. Besides, Environment settings are separated, and packages at the system level do not share environment variables that are created by \tool. Packages installed by \tool can be completely removed by deleting the root directory of the project. Therefore, \tool would not threaten the stability of Linux systems.

\section{related work}\label{sec:relwork}

\noindent \textbf{Analysis in the system ecosystems.}
Some works~\cite{wang2015graph, de2009analysis} analyzed package dependency graph on Ubuntu/Debian and present the distributions of in-degree, out-degree, betweenness centrality and clustering coefficient. Wang et al.~\cite{wang2019analysis} explored the motif evolution of Ubuntu system. Zerouali et al.~\cite{zerouali2021multi} measured multi-dimensional technical lag of packages installed in Debian-based docker images. For analysis of dependency graph in other system ecosystems, Rahkema et al.~\cite{rahkema2022analysis} provided the first study on dependency networks of package managers in iOS development and discussed the overlap, evolution and vulnerabilities in CocoaPods, Carthage, and Swift Package Manager. Gonzalez-Barahona et al.~\cite{gonzalez2009macro} performed a case study on Debian packages and present some statistical characteristics about package size, languages, and dependencies, etc. Berger et al.~\cite{berger2014variability} explored and compared the variability mechanisms in 5 ecosystems. 

In summary, works on the analysis of system ecosystems, especially Linux ecosystems focus on characteristics of the dependencies and no existing works analyze the whole ecosystem from the view of the evolutionary dependency graph cross multiple releases. We are the first to build the day-level evolutionary dependency graph with security information. Moreover, we further investigate some research questions about incompatibility issues and security issues across system releases that have not been studied in previous works.

\noindent \textbf{Analysis in the language-specific ecosystems.}
Kikas et al.~\cite{kikas2017structure} analyzed the structure and evolution of dependency networks of JavaScript, Ruby, and Rust. Various studies~\cite{liu2022demystifying, Chinthanet2021LagsIT, zimmermann2019small, decan2018impact, zerouali2022impact, chowdhury2021untriviality, latendresse2022not} have been made on vulnerability propagation and its evolution in the NPM ecosystem based on a dependency-vulnerability knowledge graph. Similarly, a lot of studies on the Java ecosystem~\cite{huang2022characterizing, wang2020empirical, wang2018dependency} and Python ecosystem~\cite{Salis2021ARP, Salis2021PyCGPC, Wang2020WatchmanMD, Mukherjee2021FixingDE} have discussed package updates, dependency conflicts, compatibility fix, call graph construction, the impact of vulnerabilities, etc. Wang et al.~\cite{Li2022NufixEF} propose NUFIX to fix the compatibility issues when the platform or dependencies are changed in .NET projects. 

Compared to analysis on the Debian ecosystems, it is more easier to construct evolutionary dependency graph for language-specific ecosystems since their historical package versions would be reserved. Linux systems of fixed release model are created branches from \verb|unstable| mirror over time. It causes another \verb|upgrade| direction that does not exist in the language-specific ecosystems.


\noindent\textbf{Package managers on Linux.}
Nowadays, some advanced package managers are proposed for Linux to overcome the shortages of default system package managers. Flatpak~\cite{flatpak:online}, Snap~\cite{snap:online}, and AppImage~\cite{url:appimage} bundle host applications with required dependencies, further provide distribution-independent packages for users. There distribution-independent packages are portable and can be run directly no matter what the Linux distribution is. Homebrew~\cite{homebrew:online} on Linux provides rolling released packages and user can take advantage of the lastet package versions on their system. Conan~\cite{url:conan} is a new package manager for \ccpp that provides duplicated packages according to compilation conditions. Nix~\cite{nix:online} is a package management tool for reproducible builds and deployments. Containers like Docker~\cite{url:docker} provide separated environments that are independent with the system. Sometimes, docker images are also Linux based and use Linux package managers to install libraries. \tool can work in Docker and improve the package management.

These package managers try to build new ecosystems rather than reuse the default ecosystems (i.e. Linux mirrors). However, it is extremely difficult to build a new repository for the whole ecosystem and ensure all packages are reliable. The repository provided by package managers should be secure, trusted, and well-maintained to prevent supply chain attacks.

\tool is a supplement tool based on existing system package managers. We use the Linux mirrors as our package repository which has the largest scale than other third-party central repositories and supports multiple architectures.

\section{Conclusion}\label{sec:conclu}
In this paper, we investigate the ecosystem gap between releases for fixed-release Linux using Debian as an example. We first build a framework to collect Debian packages and construct the evolutionary dependency graph in the Debian ecosystem. Based on the evolutionary dependency graph, we conduct an empirical study on Debian mirrors. Our study discusses compatibility issues and security threats across the Debian release and unveils many useful findings on the Debian ecosystem. To bridge the ecosystem gap between releases and improve the existing tools like APT commands on Debian, we propose a proof-of-concept package manager, \tool, that creates a separate workspace for individual projects to compensate for the disadvantage of legacy tools. We validate \tool using real-world cases that cannot be solved by legacy Debian package managers. Results show that \tool is effective to fix the incompatibilities.


\balance
\bibliographystyle{IEEEtran}
\bibliography{ref}
\balance

\end{document}